# Linked Environment Data for the Life Sciences

Maria Rüther[1,] Thomas Bandholtz[2], Antoine Logean[3]

[1]Federal Environment Agency, Corrensplatz. 1
14195 Berlin, Germany
maria.ruether@uba.de
[2]innoQ Deutschland GmbH, Halskestr. 17
40880 Ratingen, Germany
thomas.bandholtz@innoq.com
[2]innoQ Schweiz GmbH, Gewerbestr. 11
6330 Cham, Switzerland
antoine.logean@innoq.com

**Abstract.** Environment Agencies from Europe and the US are setting up a network of Linked Environment Data and are looking to crosslink it with Linked Data contributions from the life sciences.

## 1   Introduction

In 2006, Tim Berners-Lee initially formulated the principles of linking data on the Web[1]. Since then, the published Linked Data "cloud" has grown rapidly year by year. Currently, there are 216 datasets which fulfill all the requirements[2], and there are many more on their way. Of specific interest within the environmental domain is a sub-cloud from the life sciences named *Linking Open Drug Data* (LODD)[3], which gave birth to the idea of linking environment data in an international context of cooperating governmental authorities[4].

This idea was discussed in 2009 at Workshop V of the Ecoterm Group[5] with members from many European countries and the US. During 2010 the first two contributions have been published by the European Environment Agency (EEA) and several others are under development or in final test phase. The German agency will initially publish the some key instruments in the field of environmental observation that enable the long-term analysis of substance exposure of humans, species and the environment in Germany. Further on, it is envisioned to include partners from the International Environmental Specimen Bank Group (IESB).

---

[1] http://www.w3.org/DesignIssues/LinkedData.html
[2] http://lod-cloud.net/
[3] http://esw.w3.org/HCLSIG/LODD
[4] http://www.w3.org/egov/wiki/Linked_Environment_Data
[5] http://ecoterm.infointl.com

## 2 Brief Profiles of the Contributing Databases

The following profiles are based on the structure of the Comprehensive Knowledge Archive Network (CKAN)[6]. The RDF schemas mentioned in this section are listed with short descriptions in section 3.

### 2.1 Contributions of the European Environment Agency (EEA)

- General Multilingual Environment Thesaurus (GEMET)

  | | |
  |---|---|
  | *Content* | Thesaurus for terms related to the environment and environmental data in 28 languages |
  | *URL* | http://www.eionet.europa.eu/gemet |
  | *Size* | > 5,000 terms; 200,000 RDF triples |
  | *Links* | national and domain specific thesauri (planned) |
  | *Schema* | SKOS with extensions |
  | *License* | Creative commons |
  | *State* | published |

- European Nature Information System (EUNIS)

  | | |
  |---|---|
  | *Content* | species, habitats and sites across Europe |
  | *URL* | http://eunis.eea.europa.eu/ |
  | *Size* | 7,000,000 RDF triples |
  | *Links* | dbpedia, geospecies |
  | *Schema* | Darwin Core with extensions |
  | *License* | Creative commons |
  | *State* | published |

- Further Plans of the EEA

  Data on river basin districts, ground water bodies, airbase stations, NUT codes, Eurostat country codes and more.

### 2.2 Contributions of the Federal Environment Agency, Germany

After some early experiences with thesaurus-based indexing, in 2001 the Federal Environment Agency (Umweltbundesamt, UBA) started the Semantic Network Service (SNS)[7] research project. The intention was to implement the handling of vocabularies in a specialized Web service that would serve not only one information system but the environmental community in whole. SNS provides a complex, bilingual vocabulary (thesaurus, gazetteer, and chronicle) structured in form of a

---

[6] http://ckan.net/
[7] http://www.semantic-network.de

Topic Map (ISO 13250). Currently we are moving towards RDF/OWL representations. Up to now, only the thesaurus part of SNS (UMTHES) has been implemented in RDF. SNS is complemented by a species reference catalogue. The first contribution of observation data is the Environmental Specimen Bank.

- Semantic Network Service (SNS) – UMTHES® Thesaurus

  *Content*  Environmental Thesaurus in German with English translations
  *URL*      t.b.d.
  *Size*     > 50,000 terms
  *Links*    GEMET, Chronicle, Gazetteer
  *Schema*   SKOS-XL with extensions
  *License*  t.b.d.
  *State*    in final test

- Semantic Network Service (SNS) – Gazetteer

  *Content*  Named locations in Germany and their spatial intersections
  *URL*      t.b.d.
  *Size*     > 25,000 terms
  *Links*    UMTHES, Chronicle
  *Schema*   Geonames with extensions
  *License*  t.b.d.
  *State*    planned

- Semantic Network Service (SNS) – Chronicle

  *Content*  Historical and contemporary events that significantly affected the state of the environment
  *URL*      t.b.d.
  *Size*     > 1,000 events
  *Links*    UMTHES, Gazetter
  *Schema*   Linked Events Ontology with extensions
  *License*  t.b.d.
  *State*    planned

- Species Catalog

  *Content*  Organisms that have been relevant to the work of Division IV "Chemical and Biological Safety" of the UBA
  *URL*      t.b.d.
  *Size*     > 1,000 taxa
  *Links*    EUNIS, ESB
  *Schema*   SKOS, Darwin Core
  *License*  t.b.d.
  *State*    in final test

- Environmental Specimen Bank (ESB)

  | | |
  |---|---|
  | *Content* | Archive of human and environmental samples including time trends and spatial data from chemical and biological analyses since the 1980s. |
  | *URL* | http://www.umweltprobenbank.de/ |
  | *Size* | > 1.000.000 measurement data sets |
  | *Links* | EUNIS, Species Catalogue, Gazetteer, Chronicle |
  | *Schema* | SKOS, SCOVO with extensions |
  | *License* | t.b.d. |
  | *State* | in final test |

## 3 RDF Vocabularies in Use

It has been our intention to reuse existing RDF vocabularies as far as possible. The most important are listed below (in alphabetical order).

- **Darwing Core Terms**, maintained by the Taxonomy Database Working Group (TDWG)[8]. Till today, there is no canonical form of expressing Darwin Core in RDF. We have contributed our solution to the TDWG discussion as a "Simple Darwin Core" proposal (see footnote 4).
- **Geonames Ontology**[9]: add geospatial semantic information to the World Wide Web.
- **Linked Events Ontology**[10], which itself is an extension of the „An Ontology of Time for the Semantic Web"[11]. Under development.
- **SKOS(XL)**: Simple Knowledge Organization System. W3C Recommendation[12]. The XL ("extension for labels") provides means for expressing any complexity of nomenclature.
- **SCOVO**: Statistical Core Vocabulary[13]. We have proposed some extensions and specializations in order to represent the domain-specific dimensions (specimen type, analytes, sampling area, see footnote 4). In parallel, SCOVO has been further developed as the RDF Data Cube vocabulary[14], and we will synchronise with this development.

---

[8] http://www.tdwg.org/
[9] http://www.geonames.org/ontology/
[10] http://linkedevents.org/ontology
[11] http://www.w3.org/2006/time
[12] http://www.w3.org/TR/skos-reference/
[13] http://sw.joanneum.at/scovo/schema.html
[14] http://publishing-statistical-data.googlecode.com/svn/trunk/specs/src/main/html/cube.html